\documentclass[letterpaper,english,reprint, aps, pra,superscriptaddress]{revtex4-1}
\usepackage[T1]{fontenc}
\usepackage[latin9]{inputenc}
\usepackage{amsmath}
\usepackage{amssymb}
\usepackage{graphicx}
\def\oper{{\mathchoice{\rm 1\mskip-4mu l}{\rm 1\mskip-4mu l}
{\rm 1\mskip-4.5mu l}{\rm 1\mskip-5mu l}}}
\def\<{\langle}
\def\>{\rangle}
\newtheorem{Theorem}{Theorem}

\newtheorem{Proposition}{Proposition}

\usepackage{color}
\usepackage{soul}

\makeatletter

\usepackage{tikz}
\usepackage{pgfplots}
\usetikzlibrary{positioning}
\usetikzlibrary{decorations.markings}
\usetikzlibrary{arrows,decorations.text}
\usetikzlibrary{backgrounds,fit,decorations.pathreplacing}

\makeatother

\usepackage{babel}

\begin{document}

\title{A family of multipartite separability  criteria based on correlation tensor}


\author{Gniewomir Sarbicki}

\affiliation{Institute of Physics, Faculty of Physics, Astronomy and Informatics,
Nicolaus Copernicus University, Grudziadzka 5/7, 87-100 Toru\'{n},
Poland}

\author{Giovanni Scala}

\affiliation{Dipartimento Interateneo di Fisica, Universit\`a degli Studi di Bari,
I-70126 Bari, Italy}

\affiliation{INFN, Sezione di Bari, I-70125 Bari, Italy}

\author{Dariusz Chru\'{s}ci\'{n}ski}

\affiliation{Institute of Physics, Faculty of Physics, Astronomy and Informatics,
Nicolaus Copernicus University, Grudziadzka 5/7, 87-100 Toru\'{n},
Poland}

\date{\today}
\begin{abstract} A family of separability criteria based on correlation matrix (tensor) is provided.  Interestingly, it unifies several criteria known before like e.g. CCNR or realignment criterion, de Vicente criterion and derived recently separability criterion based on SIC POVMs. It should be stressed that, unlike the well known Correlation Matrix Criterion or criterion based on Local Uncertainty Relations, the new criteria are linear in the density operator and hence one may find new classes of entanglement witnesses
{and} positive maps. Interestingly, there is a natural generalization to multipartite scenario using multipartite correlation matrix. We illustrate the detection power of the above criteria on several well known examples of quantum states.

\end{abstract}



\maketitle


\section{Introduction}

Quantum entanglement is one of key features of quantum theory and provides a crucial resource for modern quantum technologies like quantum communication, quantum cryptography, and quantum calculations \cite{HHHH,QIT}. One of the tasks of the theory of quantum entanglement is to derive criteria which enables to distinguish separable and entangled states \cite{GT,HHHH}. Recall, that a state of a bipartite system living in $\mathcal{H}_A \otimes \mathcal{H}_B$ represented by a density matrix $\rho$ is separable if \cite{Werner}

\begin{equation}\label{}
  \rho = \sum_k p_k \rho^A_k \otimes \rho^B_k ,
\end{equation}
where $p_k$ is a probability distribution and $\rho^A_k$ ($\rho^B_k$) are density operators of subsytem $A$ ($B$). For low dimensional bipartite systems $2\otimes 2$ (qubit-qubit) and $2 \otimes 3$ (qubit-qutrit) this problem is completely solved by the celebrated Peres-Horodecki criterium: a state is separable if and only if it is positive partial transpose (PPT), that is, $\rho^\Gamma := ({\rm id} \otimes {\rm T})\rho \geq 0$ \cite{PPT1,PPT2}. However, for higher dimensional systems and systems composed of more than two parties the problem is notoriously difficult (actually, it belongs to the class of so called NP-hard problems \cite{NP}).

There are several separability criteria developed in the last 20 years of activity (see the reviews \cite{GT,HHHH}).
Any entangled state $\rho$ of a bipartite system can be detected a suitable entanglement witness, that is, a Hermitian operator $W$ acting in $\mathcal{H}_A \otimes \mathcal{H}_B$ such that for all separable states ${\rm Tr}(W \rho_{\rm sep}) \geq 0$ but ${\rm Tr}(W \rho) < 0$ \cite{EW,HHHH,GT,TOPICAL}.
The well known criterion based on positive maps states that $\mathcal{I}_A \otimes \Phi)\rho \geq 0$ for all positive maps $\Phi$ (it recovers PPT criterion if one takes $\Phi=T$).
{These two criteria are necessary and sufficient and related to each other via Choi-Jamio\l kowski isomorphism. While classification of entanglement witnesses (equivalently: positive maps) is not known (except the lowest dimensional cases),}
there is a number of other criteria \cite{GT,HHHH} which are not universal, i.e. do not allow to detect all entangled states, but are easily applicable and in particular allow to detect many PPT entangled states. The prominent example is realignment or computable cross-norm (CCNR) criterion \cite{R1,R2,R3}. There are also separability criteria which are nonlinear in the state of the system like for example criteria based on local uncertainty relations (LURs) \cite{LUR}, extensions of realignment criterion \cite{EXT-CCNR}  or covariance matrix criterion (CMC) \cite{COV-1,COV-2,COV-3} (see also \cite{COV-U} for the unifying approach).

In this paper we propose a unification of several bipartite separability criteria based on correlation matrix (or correlation tensor). In this category apart from CCNR one finds e.g. de Vicente criterion
{(dV)} \cite{Vicente}, separability criterion derived in \cite{Fei}  and recent criterion based on SIC POMVs (ESIC) \cite{GUHNE}. This new criterion in general is not stronger that CMC but we provide an example of PPT state which is not detected by filter{ed} CMC \cite{COV-1,COV-2} {(LFCMC)} but is detected by the new one. Our result is then generalized to multipartite scenario. We stress that the new criteria are linear in the density operator and hence  may be used to construct new classes of entanglement witness {and}
positive maps.

\section{Bipartite systems}

 Consider a bipartite system living in $\mathcal{H}_A \otimes \mathcal{H}_B$ with dimensions $d_A$ and $d_B$, respectively (in what follows we assume $d_A \leq d_B)$. Let $G^A_\alpha$ and $G^B_\beta$ denote arbitrary orthonormal basis in $\mathcal{B}(\mathcal{H}_A)$ and $\mathcal{B}(\mathcal{H}_B)$, that is, the $\<G^{A}_\mu|G^A_\nu\>_{\rm HS} = \delta_{\mu\nu}$, and the same for $G^B_\beta$ (where $\<X|Y\>_{\rm HS} := {\rm Tr}(X^\dagger Y)$ is a Hilbert-Schmidt inner product).  Now, given a bipartite state $\rho$ one defines the following correlation matrix

\begin{equation}\label{}
  C_{\alpha\beta} = \< G^A_\alpha \otimes G^B_\beta\>_{\rm HS} = {\rm Tr}(\rho  G^A_\alpha \otimes G^B_\beta) .
\end{equation}
If $\rho$ is separable, then the CCNR criterion gives the following bound for the trace norm of $C$:

\begin{equation}\label{CCNR}
  \| C\|_{\rm tr} \equiv {\rm Tr} \sqrt{CC^\dagger} \leq 1 .
\end{equation}
The norm $\|C\|_{\rm tr}$ does not depend upon the particular
{orthonormal} basis $G^A_\alpha$ and $G^B_\beta$. Let us take a particular basis consisting of Hermitian operators such that $G^A_0 = \oper_A/\sqrt{d_A}$ and $G^B_0= \oper_B/\sqrt{d_B}$ (we call it canonical basis).
It is clear that $G^A_\alpha$ and $G^B_\beta$ are traceless for $\alpha,\beta >0$. The canonical basis gives rise the following generalized Bloch representation

\begin{eqnarray}\label{}
  \rho &=& \frac{\oper_A}{d_A}  \otimes  \frac{\oper_B}{d_B} + \sum_{i>0} r^A_i G^A_i \otimes  \frac{\oper_B}{d_B} + \sum_{j>0} r^B_j  \frac{\oper_A}{d_A} \otimes G^B_j \nonumber \\
  &+& \sum_{i,j>0} t_{ij}\, G^A_i \otimes G^B_j = \sum_{\alpha=0}^{d_A^2-1}\sum_{\beta= 0}^{d_B^2-1} C^{\rm can}_{\alpha\beta} G^A_\alpha \otimes G^B_\beta   ,
\end{eqnarray}
where $r^A_i$ and $r^B_j$ are generalized Bloch vectors corresponding to reduces states $\rho_A$ and $\rho_B$, respectively, and $t_{ij}$ is usually called a correlation tensor, that is, one finds for the reduces states

$$   \rho_A = {\rm Tr}_B \rho =  \frac{\oper_A}{d_A} + \sum_{i>0} r^A_i G^A_i , $$
and

$$   \rho_B = {\rm Tr}_A \rho =  \frac{\oper_B}{d_B} + \sum_{j>0} r^B_j G^B_j . $$
We denote $C_{\alpha\beta}$ defined by the canonical basis by $C^{\rm can}_{\alpha\beta}$. Clearly $\| C^{\rm can}\|_{\rm tr} = \|C\|_{\rm tr}$. Let us introduce two square diagonal matrices:

\begin{eqnarray*}
  D^A_x = \mathrm{diag}\{x,1,\dots,1\} , \ \
  D^B_y = \mathrm{diag}\{y,1,\dots,1\} ,
\end{eqnarray*}
where $D^A_x$ is $d_A^2 \times d_A^2$ and $D^B_y$ is $d_B^2 \times d_B^2$, and the real parameters $x,y \geq 0$. Now comes the main result

\begin{Theorem} \label{TH-1} If $\rho$ is separable, then

\begin{equation}\label{xy}
  \| D^A_x C^{\rm can} D^B_y \|_{\rm tr} \leq \mathcal{N}_{A}(x) \mathcal{N}_B(y) ,
\end{equation}
where
\begin{equation}\label{NANB}
   \mathcal{N}_A(x) = \sqrt{ \frac{d_A -1 + x^2}{d_A}}\, \ \ \mathcal{N}_B(y)= \sqrt{ \frac{d_B -1 + y^2}{d_B}} ,
\end{equation}
for arbitrary $x,y \geq 0$.
\end{Theorem}
Proof: separability implies that $\rho$ is a convex combination of product states and hence
{(due to the triangle inequality for the norm)}
it is enough to check (\ref{xy}) for a product state   $\rho_A \otimes \rho_B$. One finds for the correlation matrix

$$ (C ^{\rm can})_{\alpha\beta} = R^A_\alpha R^B_\beta , $$
where $R^A_0 = {1}/{\sqrt{d_A}}$, $ R^A_i = r^A_i$ $(i \geq 1)$, and similarly for $R^B_\beta$. It implies $\|C ^{\rm can}\|_{\rm tr} = |R^A| |R^B|$, where $|R^A|^2 = \frac{1}{d_A} + |{\bf r^A}|^2$ (and the same for $R^B$). Let us observe that

$$ (D^A_x C ^{\rm can} D^B_y)_{\alpha\beta} = (R^A_x)_\alpha (R^B_y)_\beta, $$
with $R^A_x = ( x/{\sqrt{d_A}},\mathbf{r}^A)$ and $R^B_y =(y/\sqrt{d_B},\mathbf{r}^B)$. It implies

\begin{eqnarray*}
     \| D^A_x C ^{\rm can} D^B_y \|_{\rm tr} = \sqrt{\frac{x^2}{d_A} + |\mathbf{r}^A|^2}\, \sqrt{\frac{y^2}{d_B} + |\mathbf{r}^B|^2} .
\end{eqnarray*}
Finally, positivity of $\rho_A$ and $\rho_B$ requires that

$$   {\rm Tr}\rho_A^2 \leq 1 \ , \ \ {\rm Tr}\rho_B^2 \leq 1 \ , $$
which imply that the corresponding Bloch vectors $\mathbf{r}^A$ and $\mathbf{r}^B$ satisfy

$$     |\mathbf{r}^A|^2 \leq \frac{d_A-1}{d_A} \ , \ \ |\mathbf{r}^B|^2 \leq \frac{d_B-1}{d_B} , $$
and hence formula (\ref{xy}) easily follows. \hfill $\Box$

Note, that using well known inequality \cite{Horn}

\begin{equation}\label{}
  \| D^A_x C^{\rm can} D^B_y \|_{\rm tr} \leq \| D^A_x \|_\infty  \|  C^{\rm can}  \|_{\rm tr} \| D^B_y\|_\infty ,
\end{equation}
where $\|X\|_\infty = \sigma_{\rm max}(X)$ (maximal singular value of $X$), one finds for separable state $ \| D^A_x C^{\rm can} D^B_y \|_{\rm tr} \leq \| D^A_x \|_\infty \| D^B_y\|_\infty$, and hence if $x,y >1$ it implies $ \| D^A_x C^{\rm can} D^B_y \|_{\rm tr} \leq xy$. Note, however, that this condition is much weaker than (\ref{xy}).

\section{Relation to other separability criteria}

 Clearly $(x,y)=(1,1)$ reproduces CCNR criterion. Interestingly, $(x,y)=(0,0)$ reproduces separability criterion derived by de Vicente \cite{Vicente}. If $d_A=d_B$, then CCNR criterion is stronger than dV criterion. However, for bipartite states $\rho$ such that $\rho_A = \oper_A/d_A$ and $\rho_B = \oper_B/d_B$, dV criterion is stronger than CCNR  if $d_A \neq d_B$, and they are equivalent if $d_A=d_B$ \cite{Vicente}. Interestingly, we found another example of such criterion in \cite{Fei}.  After suitable renormalization the result of \cite{Fei} corresponds to $(x,y)=(\sqrt{2/d_A},\sqrt{2/d_B})$.

In a recent paper \cite{GUHNE} authors proposed an interesting separability criterion based on symmetric informationally complete positive operator valued measure (SIC POVM). Recall, that a family of $d^2$ rank-1 operators  $\Pi_i = \frac 1d |\psi_i\>\<\psi_i|$ in $d$-dimensional Hilbert space defines SIC POVM iff

$$   |\< \psi_i|\psi_j\>|^2 = \frac{d\delta_{ij}+1}{d+1} \ , \ \ \sum_{i=1}^{d^2} \Pi_i = \oper_d . $$
There is an old conjecture by Zauner that SIC POVM exists for any $d$ \cite{Zauner} (see also \cite{Renes}). So far these objects have been found for several dimensions (see \cite{SIC-rev} and \cite{SIC-2018} for the recent progress). It is, therefore, clear that the result of \cite{GUHNE} was restricted to specific dimensions only. Here we show that this criterion is universal (valid for any $d_A$ and $d_B$). Moreover, it belongs to our class (\ref{xy}) with $(x,y) = (\sqrt{d_A+1},\sqrt{d_B+1})$. The separability criterion (so called ESIC criterion) derived in \cite{GUHNE} states that if $\rho$ is separable, then

\begin{equation}\label{ESIC}
  \| P \|_{\rm tr} \leq \frac{2}{\sqrt{d_A(d_A+1)d_B(d_B+1)}} ,
\end{equation}
where $P_{\alpha\beta} = \< \Pi^A_\alpha \otimes \Pi^B_\beta\>$, and $\Pi^A_\alpha$ and $\Pi^B_\beta$ are
{elements of}
SIC POVMs in $\mathcal{H}_A$ and $\mathcal{H}_B$, respectively. It was conjectured in \cite{GUHNE} that ESIC criterion is stronger than CCNR criterion. This conjecture is supported by several examples and numerical analysis (cf. \cite{GUHNE}).

Let us observe that if $\Pi_\alpha$ define SIC POVM in $d$-dim. Hilbert space, then

\begin{equation} \label{G-Pi}
G_\alpha^{(\mp)} := \sqrt{d\left(d+1\right)}\,\Pi_\alpha - \frac{\sqrt{d+1}\mp1}{\sqrt{d^{3}}}\, \oper_d ,
\end{equation}
defines on orthonormal basis in $\mathcal{B}(\mathcal{H})$, that is, $    \< G_\alpha^{(\mp)}|G_\beta^{(\mp)} \>_{\rm HS} = \delta_{\alpha\beta}$.
Note, that this is not a canonical basis. Indeed,  $G_0^{(\mp)}$ is not proportional to $\oper_d$. However, it enjoys the following properties

\begin{equation}\label{1pm}
 {\rm Tr}\, G^{(\mp)}_\alpha = \pm \frac{1}{\sqrt{d}} \ , \ \   \sum_\alpha G^{(\mp)}_\alpha=\pm \sqrt{d} \oper_d .
\end{equation}
In what  follows we take $G_\alpha := G_\alpha^{(-)}$ (but the final result applies for $G^{(+)}_\alpha$ as well). Direct calculation shows

\begin{equation}\label{}
  \sqrt{d_A(d_A+1)d_B(d_B+1)}\, P = A C B ,
\end{equation}
where $C_{\alpha\beta}= \< G^A_\alpha \otimes G^B_\beta \>$ is a correlation matrix defined in terms of $G_\alpha = G_\alpha^{(-)}$, and

\begin{eqnarray*}\label{}
  A &=& \oper_A \otimes \oper_A + a\, \mathbb{J}_A \otimes \mathbb{J}_A , \\
  B &=& \oper_B \otimes \oper_B + b\, \mathbb{J}_B \otimes \mathbb{J}_B ,
\end{eqnarray*}
where $\mathbb{J}_A$ is $d_A \times d_A$ matrix such that $[\mathbb{J}_A]_{ij}=1$ (and similarly for $\mathbb{J}_B$). Finally

\begin{equation}\label{}
  a = \frac{\sqrt{d_A+1}-1}{d_A^2} \ , \ \ b = \frac{\sqrt{d_B+1}-1}{d_B^2} .
\end{equation}
This way we reformulated ESIC criterion (\ref{ESIC}) in terms of the correlation matrix $C_{\alpha\beta}$ as follows: if $\rho$ is separable, then

\begin{equation}\label{ACB}
  \| ACB\|_{\rm tr} \leq 2 .
\end{equation}
It should be stressed that here $C_{\alpha\beta}$ is not a canonical matrix and hence (\ref{ACB}) cannot be immediately related to (\ref{xy}). Note, however, that due to the fact that the trace norm is unitarily invariant one has

$$  \| ACB\|_{\rm tr} = \| UAU^\dagger (U CV^\dagger) V BV^\dagger \|_{\rm tr} ,  $$
for arbitrary unitary matrices $U$ and $V$. Taking $U$ and $V$ such that they diagonalize $A$ and $B$, respectively, one obtains

$$   \| ACB\|_{\rm tr} = \| D^A_x C^{\rm can} D^B_y \|_{\rm tr} , $$
with $(x,y) = (\sqrt{d_A+1},\sqrt{d_B+1})$. It proves that the original assumption about the existence of two SIC POVMs
{$\{\Pi^A_\alpha\}$ and $\{\Pi^B_\beta\}$}
is not essential and the ESIC criterion universally holds for arbitrary $d_A$ and $d_B$.

Finally, the covariance matrix criterion (CMC) \cite{COV-1,COV-2} supplemented by the procedure of local filtering (LFCMC)
 turned out to be very powerful criterion. Interestingly, for $d_A \leq d_B$ (but $d_B - d_A$ is not to big, cf. \cite{COV-2}) this criterion is equivalent to (supplemented by a local filtering) dV criterion \cite{Vicente}. Now, in our case if $\mathbf{r}^A=0$ and $\mathbf{r}^B=0$, one finds

  \begin{displaymath}
    \| D^A_x C^{\rm can} D^B_y \|_{\rm tr} = \frac{xy}{\sqrt{d_Ad_B}} +  \| D^A_0 C^{\rm can} D^B_0 \|_{\rm tr} ,
  \end{displaymath}
and hence one may wonder whether is it possible to obtain a stronger result than dV criterion. One easily finds that the function
{$\mathcal{N}_A(x) \mathcal{N}_B(y)-\frac{xy}{\sqrt{dAd_B}}$}
realizes minimum for $x \sqrt{d_B-1} = y \sqrt{d_A - 1}$ which reproduces dV \cite{Vicente}. Hence, it proves that within a class of states with maximally mixed marginals (and $d_B-d_A$ is not too big) dV condition is the strongest one.

\section{A class of entanglement witness}

Now, we show that the new separability criterion gives rise to the whole class of entanglement witnesses. Let us recall that for any $m \times n$ matrix $X$ its trace norm is given by the following formula

\begin{equation}\label{TR-norm}
  \| X \|_{\rm tr} = \max_{O \in \mathcal{O}(m,n)} \< O|X\>_{\rm HS} ,
\end{equation}
where  the maximum is performed over all isometry $m \times n$ matrices $O$. Let $\rho$ be a separable state in $\mathcal{H}_A \otimes \mathcal{H}_B$. One has therefore for any fixed $(x,y)$

$$   \|D_x^AC^{\text{can}}D_y^B \|_{\mathrm{tr}}  \leq \mathcal{N}_A(x) \mathcal{N}_B(y) $$
and hence

\begin{eqnarray}
0 &\leq& \mathcal{N}_A(x) \mathcal{N}_B(y) - \|D_x^AC^{\text{can}}D_y^B \|_{\mathrm{tr}} \nonumber\\
&= & \mathcal{N}_A(x) \mathcal{N}_B(y) \mathrm{Tr}(\rho \oper_{A} \otimes \oper_B)  \nonumber \\
&-& \max_{O \in \mathcal{O}(d^2_A,d^2_B)} \langle O | D_x^AC^{\text{can}}D_y^B  \rangle_{HS} \\
&= & \mathcal{N}_A(x) \mathcal{N}_B(y) \mathrm{Tr}(\rho \oper_{A}\otimes \oper_{B}) \nonumber \\
&+ &\min_{O \in \mathcal{O}(d^2_A,d^2_B)} \langle O | D_x^AC^{\text{can}}D_y^B  \rangle_{HS} \nonumber
\end{eqnarray}
Therefore for an arbitrary isometry $ O $

\begin{equation}\label{}
  {\rm Tr}( W^{xy}_O \, \rho) \geq 0 ,
\end{equation}
where


\begin{eqnarray}\label{eq:EW}
  W^{xy}_O =  \mathcal{N}_A(x) \mathcal{N}_B(y)\, \oper_{A}\otimes \oper_{B} +
  \sum_{\alpha,\beta} \widetilde{O}_{\alpha\beta} G_\alpha^{A}  \otimes G_\beta^B
\end{eqnarray}
and the `deformed' isometry $\widetilde{O}^{\alpha\beta}$ reads

\begin{equation*}\label{}
  \widetilde{O}^{\alpha\beta} = (D^A_{x})_{\alpha \alpha}  O^{\alpha\beta} (D^B_{y})_{\beta \beta} .
\end{equation*}
Finally, $W^{xy}_O$ has the following structure

\begin{eqnarray}\label{eq:EW-a}
  W^{xy}_O =  \sum_{\alpha,\beta} w^{\alpha\beta} G_\alpha^{A}  \otimes G_\beta^B
\end{eqnarray}
with

$$ w^{00} =  \sqrt{(d_A-1+x^2)(d_B-1+y^2)}+xyO^{00}  , $$
and

$$  w^{0\beta} = \frac{x}{\sqrt{d_A}} O^{0\beta} \ \  , \ \ w^{\alpha 0} = \frac{y}{\sqrt{d_B}} O^{\alpha 0} \ \ , \ \
w^{\alpha\beta} = O^{\alpha\beta} $$
for $\alpha,\beta > 0$. This way one obtains a big class of witnesses parameterized by $d_A^2 \times d_B^2$ isometry $O$ and two nonnegative parameters $x,y$.






\section{Multipartite criterion}

Our separability criterion (\ref{xy}) may be generalized for the multipartite scenario: consider $N$ partite system living in $\mathcal{H}_1 \otimes \ldots \otimes \mathcal{H}_N$, and let $G^{(k)}_{\alpha_k}$ denotes an orthonormal basis in $\mathcal{B}(\mathcal{H}_k)$. Given a state $\rho$ define a correlation (hyper)matrix

\begin{displaymath}
   C^N_{\alpha_1 \ldots \alpha_N} =\< G^{(1)}_{\alpha_1} \otimes \ldots  \otimes  G^{(N)}_{\alpha_N} \>_\rho.
\end{displaymath}
%
%
In order to derive generalization of (\ref{xy}), let us reformulate the definition of the trace norm (\ref{TR-norm})
\begin{equation}\label{tr-norm}
  \| X \|_{\rm tr} =   \sup_M \frac{|\langle M | X \rangle_{\rm HS}|}{\lVert M\rVert_\infty},
\end{equation}
where  the supremum is taken over all matrices of appropriate size. It is well known that supremum is always realized by some isometry (as in (\ref{TR-norm})). Now, we generalize (\ref{tr-norm}) to an arbitrary $N$-tensor $X^N_{i_1\ldots i_N}$, where

$$   \< M^N|X^N\>_{\rm HS} = \sum_{i_1,\ldots,i_N} \overline{M^N_{i_1\ldots i_N}} X^N_{i_1\ldots i_N} , $$
and the spectral (operator) norm is defined as follows

\begin{equation*}\label{sup-norm}
  \lVert M^N \rVert_\infty := \sup_{|x^{(1)}| = \dots = |x^{(N)}| = 1} | \sum_{i_1,\ldots,i_N} M^N_{i_1\dots i_N} x^{(1)}_{i_1} \dots x^{(N)}_{i_N} | .
\end{equation*}
The $N$-partite CCNR criterion reads

\begin{Proposition} If $N$-partite state is fully separable, then $\| C^N \|_{\rm tr} \leq 1 $.
\end{Proposition}
Proof: again it is enough to check it for a product state $\rho^1 \otimes \ldots \otimes \rho^N$. Since the trace norm does not depend upon the basis let us take the canonical one. One finds for the correlation hypermatrix

$$   C^N_{\alpha_1 \ldots \alpha_n} = R^1_{\alpha_1} \ldots R^N_{\alpha_N} , $$
where the vector $R^k \in \mathbb{R}^{d_k}$ reads


\begin{displaymath}
  R^k_{\alpha_k} = \< G^{(k)}_{\alpha_k} \>_{\rho^k} = {\rm Tr}( \rho^k  G^{(k)}_{\alpha_k} ) = ({1}/{\sqrt{d_k}}, \mathbf{r}^k)
\end{displaymath}
and $\mathbf{r}^k$ is a Bloch vector of $\rho^k$.
One has:
\begin{equation*}
  \langle M^N | C^N \rangle_{\rm HS} \le \lVert M^N\rVert_\infty |R^1_{\alpha_1}| \dots |R^N_{\alpha_N}|  \le \lVert M^N\rVert_\infty
\end{equation*}
and hence $\lVert C^N \rVert_{\rm tr} \le 1$. \hfill $\Box$

%
%

%
To generalize (\ref{xy}) let us define $N$ diagonal $d^2_k \times d^2_k$ matrices

$$   D^k_{x_k} = {\rm diag}\{x_k,1,\ldots,1\} ,$$
and $C^N(x_1,\ldots,x_N)$ defined as follows

$$  C^N_{i_1\dots i_N}(x_1,\ldots,x_N) =  C^N_{i_1\dots i_N} (D^1_{x_1})_{i_1i_1} \ldots (D^N_{x_N})_{i_Ni_N} . $$
One proves

\begin{Theorem} \label{TH-xxx} If $\rho$ is fully separable, then

\begin{eqnarray}\label{xxx}
   \|  C^N(x_1,\ldots,x_N) \|_{\rm tr}
    \leq \mathcal{N}_1(x_1) \ldots \mathcal{N}_N(x_N) ,
\end{eqnarray}
where
$$   \mathcal{N}_k(x_k) = \sqrt{\frac{d_k - 1 + x_k^2}{d_k}} , $$
for $k=1,\ldots,N$.
\end{Theorem}
The proof is similar to that of Theorem \ref{TH-1}. Indeed, taking again a product state $\rho^1 \otimes \ldots \otimes \rho^N$ one finds

\begin{eqnarray*}
   \< M^N|C^N(x_1,\ldots,x_N)\>_{\rm HS}
   \le \lVert M^N\rVert_\infty |R^1(x_1)| \ldots |R^N(x_N)|,
\end{eqnarray*}
where $R^k(x_k) = (x_k/\sqrt{d_k},\mathbf{r}^k)$, and hence
\begin{displaymath}
  \lVert C^N(x_1,\ldots,x_N) \rVert_{\rm tr} \le |R^1(x_1)| \ldots |R^N(x_N)| .
\end{displaymath}
Finally, note that $|R^k(x_k)|^2 = x_k^2/d_k + |\mathbf{r}^k|^2 \leq \mathcal{N}_k(x_k)$ due to $|\mathbf{r}^k|^2 \leq (d_k-1)/d_k$, which ends the proof. \hfill $\Box$

Actually, the trace-norm (or more generally Ky-Fan norm) was generalized for $N$-tensors using a procedure of so called \textit{unfoldings} \cite{MATH}: given an $X^N \in \mathbb{C}^{d_1} \otimes \ldots \otimes \mathbb{C}^{d_N}$ one defines an $n$-unfolding (or an $n$-mode  matricization of $X^N$) $X_{(n)}$ which is a $d_n \times \overline{d}_n$ matrix with $\overline{d}_n = (d_1d_2\ldots d_N)/d_n$ (see \cite{MATH} for a precise definition). Now, the Ky-Fan norm of $X^N$ is defined as follows

\begin{equation}\label{}
   \| X^N \|_{\widetilde{{\rm tr}}} := \max_{n} \| X^N_{(n)}\|_{\rm tr} .
\end{equation}
Using the same arguments one easily derives

\begin{Proposition} If $\rho$ is fully separable, then
  \begin{eqnarray}\label{KF}
   \|  C^N(x_1,\ldots,x_N) \|_{\widetilde{{\rm tr}}}
    \leq \mathcal{N}_1(x_1) \ldots \mathcal{N}_N(x_N) .
\end{eqnarray}
\end{Proposition}
Note, however that due to $\| X^N \|_{\rm tr} \leq \| X^N \|_{\widetilde{{\rm tr}}}$ the separability criterion based on (\ref{KF}) is weaker than (\ref{xxx}). The procedure of unfolding gives rise to a family of matrices each of which only controls bipartite entanglement in $d_n \times \overline{d}_n$ system. Interestingly, criterion (\ref{KF}) for $x_k=0 \ (k=1,\ldots,N)$ was already derived in \cite{Ali}, and for $x_k=\sqrt{2/d_k} \ (k=1,\ldots,N)$ it was derived in \cite{Fei}. It should be clear that if each $\mathcal{H}_k$ allows for the existence of SIC POVM, then for $x_k = \sqrt{d_k+1}$ one obtains a multipartite generalization of ESIC criterion from \cite{GUHNE}. However, as we already observed,  the existence of SICs is not essential.

\section{Detection power}

 In \cite{R3} Rudolph constructed an example of two qubit state which is entangled (and hence NPT) but it is not detected by CCNR criterion. It turns out that such state is always detected by our criterion for sufficiently big $x$ and $y$ (cf. Appendix for details). However, contrary to CMC it does not detect all NTP qubit-qubit states.

Let us consider two one-parameter families of two-qutrit states constructed from unextendable product basis (UPB) \cite{UPB-1,UPB-2}. The first family contains states of the form $\rho^{PP}_p = p \rho^{PP} + (1-p) \oper_3 \otimes \oper_3/9$, where $\rho^{PP}$ is a bound entangled state constructed by use of the Pentagon Pyramid (PP) construction. The second family contains states of the form $\rho^{Ti}_p = p \rho^{Ti} + (1-p) \oper_3 \otimes \oper_3/9$, where $\rho^{Ti}$ is a bound entangled state constructed by use of the Tiles (Ti) construction. We compare detection thresholds in this families w.r. to dV, CCNR, ESIC, and LFCMC criterion:

\begin{center}

\begin{tabular}{|c|c|c|c|c|}
  \hline
	& dV 	& CCNR 	& ESIC 	& LFCMC \\
  \hline
  PP	& .9371	& .8785	& .8739 & .8639 \\
  \hline
  Ti	& .9493 & .8897 & .8845 & .8722 \\
  \hline
\end{tabular}

\end{center}

\vspace{.1cm}

\noindent whereas our criterion detects entanglement in the PP family for $p \geq 0.8721$ ($x=y=4059.7$) and in Ti family for $p \geq 0.8822$ ($x=y= 2442.1$). Our criterion detects more than linear criteria (dV, CCNR and ESIC) but less than non-linear LFCMC.


Now, we provide an example of a qutrit-qutrit state which is detected neither by CCNR nor by ESIC  but it is detected by (\ref{xy}).
Consider a chessboard state \cite{Bruss}  defined in terms of four orthogonal vectors in $\mathbb{C}^3\otimes\mathbb{C}^3$:
  \begin{align*}
  &  |V_1\rangle = |m,0,s;0,n,0;0,0,0\rangle \\
  &  |V_2\rangle = |0,a,0;b,0,c;0,0,0\rangle \\
  &  |V_3\rangle = |n^*,0,0;0,-m^*,0;t,0,0\rangle \\
  &  |V_4\rangle = |0,-b^*,0;a^*,0,0;0,d,0\rangle
  \end{align*}
giving rise to $  \rho  = \mathcal{N}\sum_i |V_i\rangle\langle V_i|$,
with $\mathcal{N}$ being a normalization factor.
Let us consider the  mixture with white noise $\rho_p = p \rho + (1-p) \oper_3 \otimes \oper_3/9$. It is shown in the Appendix that by taking a suitable parameters we may construct a PPT state $\rho_p$ that is detected neither by CCNR nor by ESIC, nor by filter CMC \cite{GUHNE} but it is detected by (\ref{xy}) for $(x,y)=(5.5,5.9)$ (cf. the Figure).
 \begin{center}
\begin{figure}
  \centering
    \includegraphics[width=.40\textwidth]{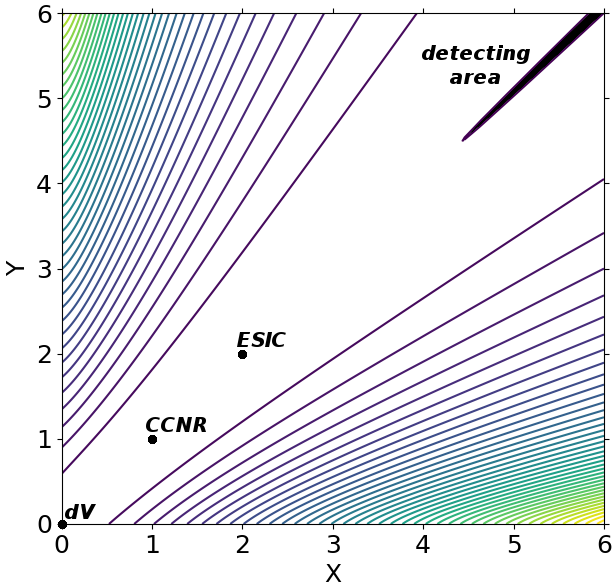}
\caption{Contour levels of the function $f(x,y) =\mathcal{N}_{A}(x) \mathcal{N}_B(y) - \| D^A_x C^{\rm can} D^B_y \|_{\rm tr}$. In the \emph{detecting area} $f(x,y)<0$, since the state is detected according to our criterion (\ref{xy}) with parameters $(x,y)$ as in Eq. \eqref{NANB}. Three characteristic points on the $xy$ plane which restore well-known criteria: $(0,0)$ -- dV; $(1,1)$ -- CCNR, and $(2,2)$ -- ESIC. }
\end{figure}
  \end{center}

\section{Conclusions}

In this paper we provide a new family of separability criteria. Interestingly, it unifies several criteria known before like e.g. CCNR or realignment criterion, de Vicente criterion and derived recently separability criterion based on SIC POVMs. All these criteria are based on the universal object --  correlation matrix defined in terms of Hermitian orthonormal basis in the operator space. It should be stressed that, unlike the well known CMC or LUR, the new criteria are linear in the density operator. This property enables us to provide new classes of entanglement witnesses {and} positive maps. Interestingly, there is a natural generalization to multipartite scenario using multipartite correlation matrix and
{multipartite generalizations of matrix norms}. This approach generalizing multipartite setting analyzed in \cite{Fei,Ali,Vicente3} and it is essentially different from original CCNR proposed to multipartite setting in \cite{Chen-multi}. It would be interesting to compare two multipartite criteria based on (\ref{xxx}) and (\ref{KF}) analyzing various multipartite states \cite{GT,B1,B2}. Another interesting point would be to analyze which entanglement witnesses corresponding to the above separability criteria are optimal \cite{OPT}.

\begin{acknowledgments}
DC and GSa were supported by the Polish National Science
Centre project 2015/19/B/ST1/03095.  GSc  thanks
S. Pascazio, P. Facchi and F. V. Pepe for invaluable human and scientific
support, for suggestions and encouragements which led to the realization
of the present work.
\end{acknowledgments}

\appendix

\begin{widetext}

\section{Two qubit example of Rudolph \cite{R3} }

Consider a qubit-qubit density operator

\begin{equation}\label{rho-R}
  \rho = \frac 12 \left( \begin{array}{cccc}
                  1+r & 0 & 0 & t \\
                  0 & 0 & 0 & 0 \\
                  0 & 0 & s-r & 0 \\
                  t & 0 & 0 & 1-s
                \end{array} \right),
\end{equation}
where the real parameter $\{r,s,t\}$ are taken such that $\rho \geq 0$. This state is NPT (entangled) iff $|t|> 0$. One finds for the correlation matrix

\begin{equation}\label{}
  C^{\rm can} = \frac 12 \left( \begin{array}{cccc}
                  1 & 0 & 0 & r \\
                  0 & t & 0 & 0 \\
                  0 & 0 & -t & 0 \\
                  s & 0 & 0 & 1+r-s
                \end{array} \right),
\end{equation}
and hence $\| C^{\rm can}\|_{\rm tr} = |t| + g(r,s)$ \cite{R3}, and in general $\| C^{\rm can}\|_{\rm tr} \leq 1$ even if $|t|>0$.

Now, using our criterion one finds

\begin{equation}\label{}
  D^A_x C^{\rm can} D^B_y = \frac 12 \left( \begin{array}{cccc}
                  xy & 0 & 0 & x r \\
                  0 & t & 0 & 0 \\
                  0 & 0 & -t & 0 \\
                  y s & 0 & 0 & 1+r-s
                \end{array} \right),
\end{equation}
and hence for a separable (PPT) state

$$ \|  D^A_x C^{\rm can} D^B_y \|_{\rm tr} = |t| + f(x,y;r,s) , $$

$$  f(x,y,r,s) = \sqrt{\lambda_+(x,y;r,s)} + \sqrt{\lambda_-(x,y;r,s)} , $$

\begin{eqnarray*}
    \lambda_\pm(x,y;r,s) &=& \frac 18 \Big( (1+r-s)^2 + s^2 x^2 + r^2 y^2 + x^2y^2 \\
  &\pm&  \sqrt{ ((1+r-s)^2 + s^2 x^2 + r^2 y^2 + x^2y^2)^2 - 4 (1+r)^2(1-s)^2x^2y^2 } \Big) .
\end{eqnarray*}
Note, that in the limit $x,y \to \infty$

$$ \lambda_+(x,y;r,s) \to \frac{x^2y^2}{4}  \ , \ \ \lambda_-(x,y;r,s) \to 0 \ ,\  \ f(x,y;r,s) \to \frac{xy}{2} , $$
and hence for PPT (separable) state our criterion

$$   |t| + f(x,y;r,s)  \leq \sqrt{ \frac{1+x^2}{2}} \sqrt{\frac{1+y^2}{2}} , $$
gives in the limit $x,y\to \infty $ the condition $ |t| \leq 0 $
which recovers PPT condition for (\ref{rho-R}).

\section{Chessboard state \cite{Bruss}}

Taking the following parameters
  \begin{align*}
    & a=0.3346 &&
    & b=-0.1090 &&
    & c=-0.6456 \\
    & d=0.8560 &&
    & m=0.4690 &&
    & n=-0.3161 \\
    & s=-1.0178 &&
    & t=-0.6085 &&
    & p=0.8062
  \end{align*}
  one obtains the following PPT density matrix (whose entanglement is not detected by realignment, CMC criterion and ESIC criterion):
    \begin{equation}
      \rho_{p} = \left[ \begin{array}{ccc|ccc|ccc}
	0.0964 & 0 & -0.1118 & 0 & 0 & 0 & 0.0450 & 0 & 0\\
	0 & 0.0505 & 0 & 0 & 0 & -0.0506 & 0 & -0.0218 & 0\\
	-0.1118 & 0 & 0.2641 & 0 & 0.0753 & 0 & 0 & 0 & 0\\
	\hline
	0 & 0 & 0 & 0.0505 & 0 & 0.0165 & 0 & -0.0671 & 0\\
	0 & 0 & 0.0753 & 0 & 0.0964 & 0 & 0.0668 & 0 & 0\\
	0 & -0.0506 & 0 & 0.0165 & 0 & 0.1191 & 0 & 0 & 0\\
	\hline
	0.0450 & 0 & 0 & 0 & 0.0668 & 0 & 0.1082 & 0 & 0\\
	0 & -0.0218 & 0 & -0.0671 & 0 & 0 & 0 & 0.1931 & 0\\
	0 & 0 & 0 & 0 & 0 & 0 & 0 & 0 & 0.0215
      \end{array} \right]
    \end{equation}
We calculate the quantity:
  \begin{align}
    \sqrt{\frac{2+x^2}{3}}\sqrt{\frac{2+y^2}{3}}
    - \lVert D^A_x C^{\rm can} D^B_y \rVert_{\mathrm{tr}}
  \end{align}
  for $(x,y)=(5.8,5.9)$. It should be nonegative for separable states. We get $\approx -5.45 \times 10^{-5}$ and hence we detect entanglement in the state.   On the other hand performing a local filtering of $\rho_p$:
  \begin{equation}
    \rho_{\rm LF} = \frac{A\otimes B \rho_p A^\dagger\otimes B^\dagger}{\mathrm{Tr}(A\otimes B \rho_p A^\dagger\otimes B^\dagger)}
  \end{equation}
  with operators:
  \begin{align}
    A = \left[ \begin{array}{ccc} 1.2970 & 0 & -0.0770 \\ 0 & 1.4374 & 0 \\ -0.0892 & 0. & 1.2698 \end{array} \right] \ , \ \
    B = \left[ \begin{array}{ccc} 0.9171 & 0 & 0.1126 \\ 0 & 0.7412 & 0 \\ 0.1126 & 0 & 0.6961 \end{array} \right]
  \end{align}
  one obtains a state $\rho_{\rm LF}$ with maximally mixed partial traces:
    \begin{equation}
      \rho_{\rm LF} = \left[ \begin{array}{ccc|ccc|ccc}
	0.0962 & 0 & -0.0717 & 0 & 0.0067 & 0 & 0.0466 & 0 & 0.0113\\
	0 & 0.0497 & 0 & -0.0028 & 0 & -0.0480 & 0 & -0.0334 & 0\\
	-0.0717 & 0 & 0.1878 & 0 & 0.0718 & 0 & 0.0113 & 0 & -0.0131\\
	\hline
	0 & -0.0028 & 0 & 0.0980 & 0 & 0.0522 & 0 & -0.0827 & 0\\
	0.0067 & 0 & 0.0718 & 0 & 0.1095 & 0 & 0.0821 & 0 & 0.0052\\
	0 & -0.0480 & 0 & 0.0522 & 0 & 0.1259 & 0 & -0.0069 & 0\\
	\hline
	0.0466 & 0 & 0.0113 & 0 & 0.0821 & 0 & 0.1391 & 0 & 0.0194\\
	0 & -0.0334 & 0 & -0.0827 & 0 & -0.0069 & 0 & 0.1740 & 0\\
	0.0113 & 0 & -0.0131 & 0 & 0.0052 & 0 & 0.0194 & 0 & 0.0198
      \end{array} \right]
    \end{equation}
Calculating quantity:
  \begin{equation}
    \sqrt{\frac{2}{3}}\sqrt{\frac{2}{3} }
      - \lVert D^A_0 C^{\rm can} D^B_0 \rVert_{\mathrm{tr}}
  \end{equation}
one gets $\approx 5.41 \times 10^{-3}$, hence the state is not detected by the Covariance Matrix Criterion after local filtering making its partial traces maximally mixed.

\end{widetext}


%
%
%
%
%

\end{document}